\documentclass[pra,showpacs,showkeys,amsfonts,amsmath, twocolumn]{revtex4}
\usepackage{bm}
\usepackage{graphicx}
\RequirePackage{mathptm}

\numberwithin{equation}{section}

\newcommand{\ro}{\rho}

\newcommand{\al}{\alpha}
\newcommand{\be}{\beta}
\newcommand{\el}{\ell}
\newcommand{\g}{\gamma}

\newcommand{\Ga}{\Gamma}

\newcommand{\I}{\openone}
\newcommand{\ket}[1]{|{#1}\rangle}
\newcommand{\bra}[1]{\langle {#1} |}
\newcommand{\cH}{{\mathcal H}}
\newcommand{\Z}{\mathbb Z}
\newcommand{\C}{\mathbb C}

\newcommand{\M}{\mathbb M}
\newcommand{\PP}{\mathbb P}
\newcommand{\Wu}{\mathfrak W}

\newcommand{\tr}{\mathrm{tr}\,}

\begin{document}
\title{Entanglement of a class of mixed two-qutrit states}
\author{{\L}ukasz Derkacz}
\affiliation{Institute of Theoretical Physics\\ University of
Wroc{\l}aw\\
Pl. M. Borna 9, 50-204 Wroc{\l}aw, Poland}
\author{Lech Jak{\'o}bczyk\footnote{
E-mail addres: ljak@ift.uni.wroc.pl}} \affiliation{Institute of
Theoretical Physics\\ University of
Wroc{\l}aw\\
Pl. M. Borna 9, 50-204 Wroc{\l}aw, Poland}
\begin{abstract}
We compute the measure of entanglement for some classes of states
belonging to the simplex of Bell-diagonal states of two qutrits.
\end{abstract}
\pacs{03.67.Mn, 03.67.Hk} \keywords{entanglement; two qutrits;
negativity} \maketitle
\section{Introduction}
Description of a detailed structure of the set of states of quantum
system is one of the most important problems in quantum theory. From
the general point of view, the state of the system can be considered
as the probability measure $m$ on the lattice of projectors
$\PP(\cH)$ defined on the Hilbert space $\cH$ of the system. By the
Gleason theorem \cite{Gl}, such a measure can be characterized by
the trace class nonnegative operator $\ro$ satisfying $\tr\, \ro=1$,
and
$$
m(P)=\tr (\ro P),\quad P\in \PP(\cH)
$$
Thus, the set of states can be identified with convex set of all
nonnegative operators with trace equal to one, with extremal points
corresponding to pure vector states. The convex structure of the set
of states can be exploited to get some knowledge about the quantum
states (see e.g. \cite{BZ}), but the structure of this set, even in
the case of $n$ - level quantum system, is completely known only for
$n=2$.
\par
The simplest bipartite quantum system in which the phenomenon of
entanglement occurs consist of two two - level systems (qubits). The
structure of resulting four - level quantum system was the subject
of many studies, specially in the context of characterizing of
entanglement and entanglement measures (see e.g. \cite{HHH, HW, W})
and the geometry of quantum states \cite{JS,K,S}. The important
problem of evolution of entanglement in realistic quantum systems
interacting with their environments, is also  modeled by the
dynamics of two - qubit states. In that case the processes of
degradation or creation of entanglement can be studied in details
(see e.g. \cite{YE, Jm, J, FT,TF}) by using analytically computable
measure of entanglement given by concurrence \cite{HW,W}.
\par
Much more complex and  interesting are processes of disentanglement
or creation of entanglement involving multilevel atoms. In such
cases, quantum interference between different radiative transitions
can influence the dynamics of the system. For a pair of three -
level atoms the role of interference was studied in Ref. \cite{DJ}.
It is obvious that  detailed knowledge of the set of states is
crucial for systematic analysis of such phenomena. Even when we
focus on two three - level systems (two qutrits), analytic
description of entanglement of general states is not possible, so we
can try to solve this problem for specific classes of states. In
this context, the interesting study of  two - qutrit states appeared
recently in \cite{BHN}, where the analog of the set of Bell -
diagonal states for two qubits - the simplex $\Wu$, was
investigated. Since $\Wu$ lives in nine - dimensional real linear
space (instead of eighty - dimensional space for all two - qutrit
states), the investigation of its properties is simpler and one can
use symmetries and equivalences inside $\Wu$ to classify interesting
quantum states and discuss the geometry of entanglement.
\par
In the present paper we also study the properties of the simplex
$\Wu$, but we concentrate on the computation of the measure of
entanglement, given by negativity, for some classes of states in
$\Wu$. Using the symmetry of $\Wu$ we can classify mixed states
inside $\Wu$ with respect to their local equivalence. As follows
from \cite{BHN}, all mixtures of two Bell states (for fixed mixing
probabilities) are equivalent and there are two classes of mixtures
of three Bell states and two of four states. For those states we
compute its negativity analytically or numerically.  To obtain some
information about general mixed states from $\Wu$, we consider two
parameters: negativity and degree of mixture given by linear
entropy. On the entropy - negativity plane, the simplex $\Wu$ is
represented by the region bounded by two curves. It turns out that
on the upper curve lie all Werner states i.e. mixtures of maximally
mixed state and any maximally entangled Bell state. It means that
within $\Wu$, Werner states have maximal allowed negativity for
given linear entropy. By numerical generation of states lying above
the curve of Werner states we show, that similarly as in the case of
two qubits, Werner states do not maximize entanglement for given
entropy.
\section{A class of two-qutrit states}
We start with the construction of simplex $\Wu$. Let us fix the
basis $\ket{0},\; \ket{1},\; \ket{2}$ for one-qutrit space $\C^{3}$.
In the space of two qutrits $\C^{3}\otimes \C^{3}$, consider the
maximally entangled pure state of the form
\begin{equation}
\Psi_{0,0}=\frac{1}{\sqrt{3}}\sum\limits_{k=0}^{2}\ket{k}\otimes
\ket{k}\label{psi00}
\end{equation}
To construct the basis of $\C^{3}\otimes \C^{3}$ consisting of
maximally entangled pure "Bell-like" states, we can proceed as
follows \cite{BHN}. Let $\M$  be the set of pairs of indices
$(m,n)$, where $m,n \in \Z_{3}$ i.e. the addition and multiplication
of indices are understood as modulo $3$. For each $\al=(m,n)\in \M$,
define the unitary operator $W_{\al}$ as
\begin{equation}
W_{\al}=W_{(m,n)}=\sum\limits_{k=0}^{2}e^{\frac{2\pi
i}{3}kn}\ket{k}\bra{k+m}\label{W}
\end{equation}
One can check that the operators $W_{\al}$ have the following
properties:
\begin{eqnarray*}
&&W_{(0,0)}=\I\\[2mm]
&&W_{(m,n)}^{\dagger}=e^{\frac{2\pi i}{3}nm}\,W_{(-m,-n)}\\[2mm]
&&W_{(m,n)}W_{(k,l)}=e^{\frac{2\pi i}{3}nk}\,W_{(m+k,n+l)}
\end{eqnarray*}
Then to each point $\al\in\M$ we associate the vector $\Psi_{\al}\in
\C^{3}\otimes\C^{3}$, defined as
\begin{equation}
\Psi_{\al}=(W_{\al}\otimes \I)\, \Psi_{0,0}\label{psialfa}
\end{equation}
In this way we obtain nine Bell-like maximally entangled vectors,
which form a basis of the space $\C^{3}\otimes \C^{3}$. Notice that
all states (\ref{psialfa}) are locally equivalent.
\par
The class of states of two qutrits we are considering in the paper,
contains all Bell-diagonal states i.e. the mixtures of pure states
$\Psi_{\al},\; \al \in \M$:
\begin{equation}
\Wu=\{\sum\limits_{\al\in \M}p_{\al}P_{\al}\,:\, p_{\al}\geq 0,\;
\sum\limits_{\al}p_{\al}=1\}\label{sympleks}
\end{equation}
where $P_{\al}=\ket{\Psi_{\al}}\bra{\Psi_{\al}}$. Although the
simplex (\ref{sympleks}) constitutes only a small part of the set of
all states, it contains interesting entangled mixed states and the
problem of separability of general state from $\Wu$ is still
involved. To proceed further, we  study some equivalence classes
within $\Wu$, with respect to the local operations which do not
change entanglement properties of states. As was shown in Ref.
\cite{BHN}, such mappings can be considered as transformations of
the set of pairs of indices, which form a group of affine
transformations
\begin{equation}
\begin{pmatrix}k\\l\end{pmatrix}\mapsto
\begin{pmatrix}m&n\\p&q\end{pmatrix}\begin{pmatrix}k\\l\end{pmatrix}+\begin{pmatrix}j\\r\end{pmatrix}
\label{grupa}
\end{equation}
where all elements are from $\Z_{3}$ and $mp-pn\neq 0$.
\par
Using the transformations (\ref{grupa}), all subsets of $\M$ can be
classified with respect to the equivalence relation. In particular,
there is one class of single points, one of two points, two classes
of three points and two of four points. For three points, all lines
in $\M$ i.e. sets of the form
$$
\el=\{(j,k),\; (j+n,k+m),\; (j+2n,k+2m)\}
$$
are equivalent. All other sets of three points form another
equivalence class. In the case of four points, all sets
$$
\Gamma =\el \cup \{\al\},\quad\al \notin \el
$$
where $\el$ is any line, are equivalent. The second class contain
all "rectangles" $Q$ in $\M$ defined as
$$
Q=\{ (j,k),\; (j+n,k),\; (j+n,k+m),\; (j,k+m)\}
$$
To the equivalence classes of points from $\M$ correspond
equivalence classes of states from  the simplex $\Wu$. In
particular, single points $\al$ define pure states $P_{\al}$ which
are locally equivalent by definition. From the pair of points
$\{\al, \be\}$ we obtain mixtures
\begin{equation}
\ro_{\{\al,\be\}}=p_{\al}P_{\al}+p_{\be}P_{\be}\label{ro2}
\end{equation}
and for fixed $p_{\al},\; p_{\be}$ all such states are locally
equivalent. Three points $\{\al,\; \be,\; \g \}$ not lying on any
line define a class of equivalent mixed states
\begin{equation}
\ro_{\{\al,\;\be,\; \g \}}=p_{\al}
P_{\al}+p_{\be}P_{\be}+p_{\g}P_{\g}\label{ro3}
\end{equation}
and the second class is given by the states
\begin{equation}
\ro_{\el}=\sum\limits_{\al\in\el}p_{\al}P_{\al}\label{rol}
\end{equation}
for any line $\el$ in $\M$. Finally, there are two classes of states
corresponding to four points:
\begin{equation}
\ro_{Q}=\sum\limits_{\al\in Q}p_{\al}P_{\al}\label{roQ}
\end{equation}
for any rectangle $Q$, and
\begin{equation}
\ro_{\Gamma}=\sum\limits_{\al\in\Ga}p_{\al}P_{\al}\label{roG}
\end{equation}
for any set $\Ga=\el\cup \{\al\}$.
\section{NPPT states and negativity}
In the case of compound quantum systems, the main problem is how to
distinguish between separable and entangled states. Pure states of
the system are entangled if the parts of the system do not have pure
states of their own, but for mixed states one cannot deduce if they
are entangled by considering partial states. For two qubits there is
a simple necessary and sufficient criterion for entanglement: states
which are non positive after partial transposition (NPPT states) are
entangled \cite{P,HHH}. In the case of two qutrits, we only know
that all NPPT states are entangled, but there are entangled states
which are positive after this operation \cite{H}. For NPPT states,
the natural measure of entanglement is based on the trace norm of
the partial transposition $\ro^{PT}$ of the state \cite{VW}. One
defines \textit{negativity} of the state $\ro$ as
\begin{equation}
N(\ro)=\frac{||\ro^{PT}||_{1}-1}{2}\label{neg}
\end{equation}
$N(\ro)$ is equal to the absolute value of the sum of negative
eigenvalues of $\ro^{PT}$ and is an entanglement monotone \cite{VW}.
Notice that (\ref{neg}) is normalized such that $N(\ro)=1$ if and
only if $\ro$ is maximally entangled.
\subsection{Negativity for $\ro\in \Wu$}
\begin{figure}[b]
\centering {\includegraphics[height=40mm]{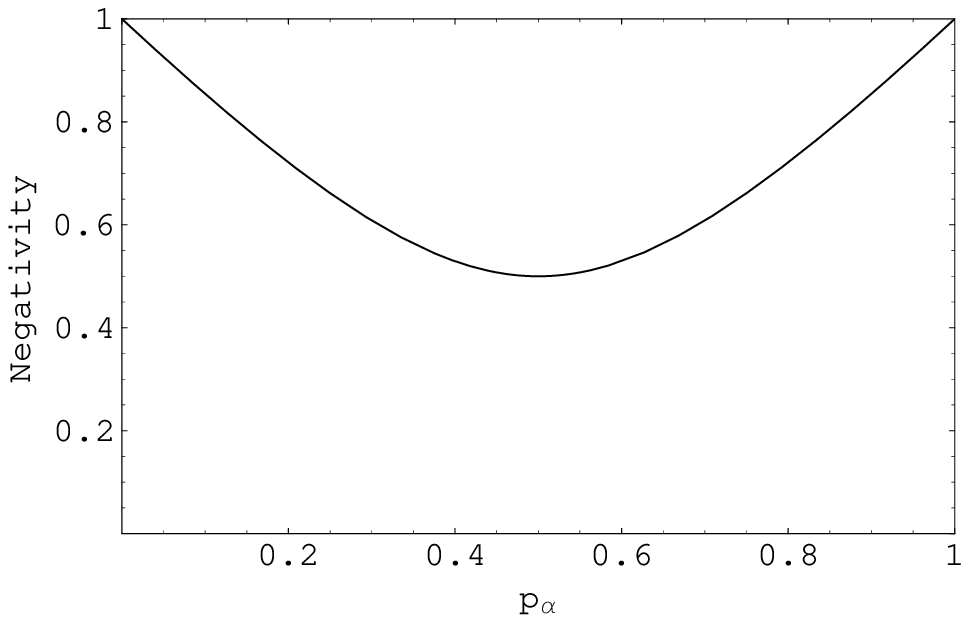}}\caption{Plot of
$N(\ro_{\{\al,\be\}})$}
\end{figure}
 To get some insight into the properties of this measure of
entanglement, we compute it  analytically or numerically for some
states from the simplex $\Wu$.
\par
 For any state (\ref{ro2}) corresponding to the pair of points in
 $\M$, there is a simple formula for negativity
\begin{equation}
N(\ro_{\{\al,\be\}})=\sqrt{1-3p_{\al}p_{\be}},\quad
p_{\al}+p_{\be}=1\label{neg2}
\end{equation}
Notice that the states (\ref{ro2}) are always entangled and minimal
value of negativity is attained for symmetric combination of pure
states (FIG.1). For three points lying on some line $\el$, one can
prove by direct computation that
\begin{equation}
N(\ro_{\el})=\sqrt{\frac{1}{2}\sum\limits_{\al,\be\in\el\atop\al\neq
\be}(p_{\al}-p_{\be})^{2}}\label{negl}
\end{equation}
In that case, symmetric combination of pure states i.e. the state
(\ref{rol}) such that $p_{\al}=p_{\be}=p_{\g}=1/3$ produces
separable state. For the remaining states belonging to other class
(\ref{ro3}), we can compute its negativity  numerically and we get
$$
N(\ro_{\{\al,\be,\g\}})\geq N(\ro_{\el})
$$
for every state $\ro_{\{\al,\be,\g\}}$ given by convex combination
(\ref{ro3}), where $\al,\be,\g$ do not belong to the line $\el$ but
the coefficients $p_{\al},p_{\be},p_{\g}$ are the same as in the
decomposition (\ref{rol}) of $\ro_{\el}$. Moreover,
$N(\ro_{\{\al,\be,\g\}})$ is always greater then zero (FIG.2).
\begin{figure}[h]
\centering {\includegraphics[height=50mm]{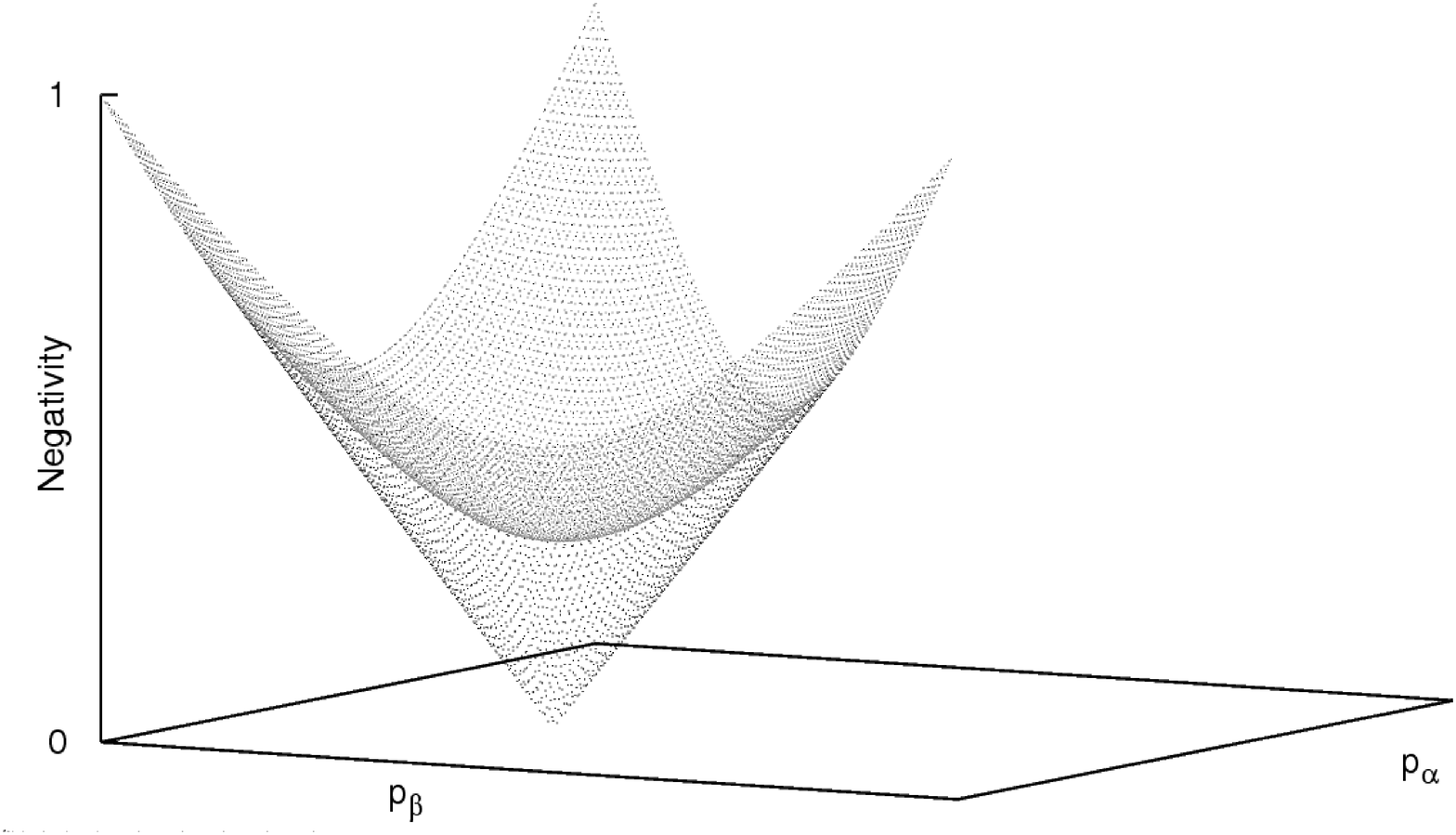}}\caption{Plot of
$N(\ro_{\el})$ (light gray surface) and $N(\ro_{\{\al,\be,\g\}})$
(gray surface)}
\end{figure}
\par
For nonequivalent states corresponding to  four points in $\M$, we
are only able to compute its negativity numerically. In particular,
we study the states
\begin{equation}
\ro_{Q}=p_{\al}P_{(00)}+p_{\be}P_{(10)}+p_{\g}P_{(11)}+p_{\delta}P_{(01)}\label{roQ3}
\end{equation}
and
\begin{equation}
\ro_{\Gamma}=p_{\al}P_{(00)}+p_{\be}P_{(10)}+p_{\g}P_{(20)}+p_{\delta}P_{(21)}\label{roG3}
\end{equation}
We check numerically that
$$
N(\ro_{Q})\geq N(\ro_{\Gamma})
$$
for all such states. We also plot the surfaces of negativity for
fixed $p_{\g}$ (see FIG.3).

\begin{figure}[h]
\centering
{\includegraphics[height=40mm]{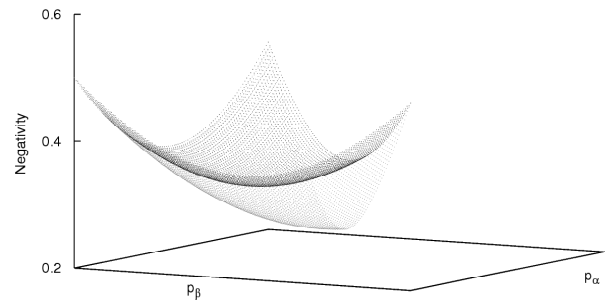}}\caption{Negativity for
$p_{\g}=0.5$: gray surface - the states (\ref{roQ3}), light gray
surface - the states (\ref{roG3}).}
\end{figure}
\subsection{Negativity versus entropy}
General mixed states from the simplex $\Wu$ can be to some extend
characterized by two parameters: degree of entanglement and mixture.
To quantify the first we take negativity, and the second is given by
normalized linear entropy
\begin{equation}
S_{L}(\ro)=\frac{9}{8}\;\tr (\ro-\ro^{2})\label{lentropy}
\end{equation}
which vanishes for all pure states and equals to $1$ for maximally
mixed state
\begin{equation}
\ro_{\infty}=\frac{1}{9}\I_{9}
\end{equation}
One can find general bound on (\ref{lentropy}) for states which are
positive after partial transposition (PPT states), by applying the
result: for $N\times N$ bipartite quantum system if $\ro$ satisfies
\begin{equation}
\tr\,\ro^{2}\leq \frac{1}{N^{2}-1}\label{purity}
\end{equation}
then $\ro$ is PPT state \cite{ZHSL}. Thus for two-qutrits we obtain
the bound: if $ S_{L}(\ro)\geq \frac{63}{64} $ then $\ro$ is PPT
state and NPPT states can only have linear entropies less than this
number.
\par
It would be interesting to find the physically allowed degree of
entanglement and mixture for NPPT states of two qutrits. In
two-qubit case, the subset of the entropy - concurrence plane
corresponding to possible physical states was characterized by Munro
et al. \cite{M}. In particular, the states lying on the boundary of
this set were identified with maximally entangled mixed states.
Similar characterization for two qutrits (in terms of entropy and
negativity) is not easy and we restrict our investigations to the
case of states $\ro\in \Wu$. Since these states are diagonal in the
basis (\ref{psialfa})
\begin{equation}
S_{L}(\ro)=\frac{9}{8}\;\left(1-\sum\limits_{\al\in\M}p_{\al}^{2}\right)\label{lentrW}
\end{equation}
\begin{figure}[h]
\centering
{\includegraphics[height=95mm,angle=270]{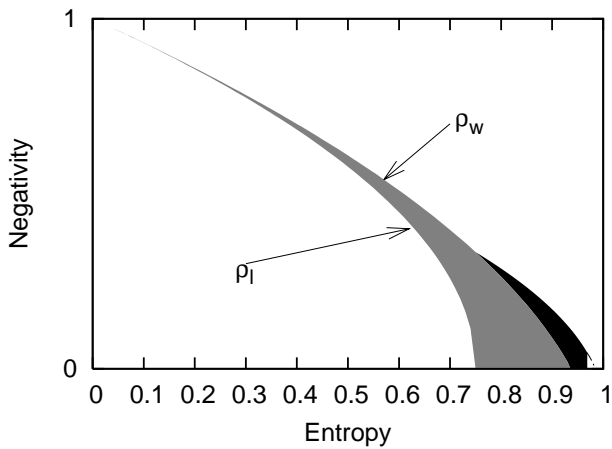}}\caption{The set
$(S_{L},N)$ for the states $\ro\in\Wu$ (gray region) with the curves
(\ref{Kl}) and (\ref{KW}) indicated. The dark region corresponds to
numerically generated two-qutrit states. }
\end{figure}
Explicit relation between linear entropy and negativity can be
established for the class of states (\ref{rol}).  Combining
(\ref{negl}) with (\ref{lentrW}), one finds that all states
$\ro_{\el}$ lie on the curve on the entropy - negativity plane,
given by the equation
\begin{equation}
n=\sqrt{1-\frac{4}{3}s},\quad s\in [0,3/4]\label{Kl}
\end{equation}
For the Werner states
\begin{equation}
\ro_{W}=(1-p)\ro_{\infty}+pP_{\al},\quad \al\in\M,\; p\in
[0,1]\label{Werner}
\end{equation}
direct calculations show that
\begin{equation}
N(\ro_{W})=\frac{1}{3}(4p-1),\quad p>\frac{1}{4}\label{Wneg}
\end{equation}
and
\begin{equation}
S_{L}(\ro_{W})=1-p^{2}\label{Wentr}
\end{equation}
thus all states (\ref{Werner}) lie on the curve
\begin{equation}
n=\frac{1}{3}\left(4\sqrt{1-s}-1\right),\quad s\in
[0,15/16]\label{KW}
\end{equation}
It turns out that the curves (\ref{Kl}) and (\ref{KW}) form the
boundary of the region on the entropy - negativity plane that is
occupied by the states $\ro\in\Wu$ (FIG.4). It means that in the
class $\Wu$, Werner states have maximal allowed negativity for a
given entropy, whereas the states $\ro_{\el}$ have a minimal
negativity under the same conditions. It is obvious that the whole
region below the curve (\ref{Kl}), corresponds to physical states of
two qutrits. On the other hand, since the maximal allowed linear
entropy of Werner states is less than the value which follows from
the bound (\ref{purity}), there should exist physical states above
the curve (\ref{KW}). In the other words, the Werner states are not
maximally entangled states for given linear entropy. This conjecture
was confirmed by numerical generation of two-qutrit states lying
above the curve of Werner states (FIG.4). It is  difficult to
generate such states, and we were able to find them only for limited
values of negativity. For this reason, we cannot predict the curve
of maximally entangled mixed states.

\end{document}